\documentclass[aps,prl,twocolumn,superscriptaddress,showpacs]{revtex4}
\usepackage{graphicx}

\begin{document}

\noindent \textbf{Comment on ``Limits on the Time Variation of the
  Electromagnetic Fine-Structure Constant...''}

\smallskip

In their Letter \cite{SrianandR_04a} (also \cite{ChandH_04a}),
Srianand {\it et al.} analysed optical spectra of heavy-elements in 23
absorbers along background quasar sight-lines, reporting limits on
variations in the fine-structure constant, $\alpha$:
$\Delta\alpha/\alpha=(-0.06\pm0.06)\times10^{-5}$.  This would
contradict previous evidence \cite[e.g.][]{WebbJ_99a,MurphyM_03a} for
a smaller $\alpha$ in the absorption clouds compared to the
laboratory: $\Delta\alpha/\alpha=(-0.57\pm0.11)\times10^{-5}$
\cite{MurphyM_04a}.  Here we demonstrate basic flaws in the analysis
of \cite{SrianandR_04a} using the same data and absorption profile
fits.

For each absorber, $\Delta\alpha/\alpha$ is measured using a $\chi^2$
minimization of a multiple-component Voigt profile fit to the
absorption profiles of several transitions. The column densities,
Doppler widths and redshifts defining the components are varied
iteratively until the decrease in $\chi^2$ between iterations falls
below a specified tolerance, $\Delta\chi^2_{\rm tol}$. In our
approach, we simply add $\Delta\alpha/\alpha$ as an additional free
parameter whereas \cite{SrianandR_04a} keep it as an external one: for
each fixed input value of $\Delta\alpha/\alpha$ the other, free
parameters are varied to minimize $\chi^2$. The functional form of
$\chi^2$ implies that, in the vicinity of the best-fitting
$\Delta\alpha/\alpha$, the `$\chi^2$ curve' -- the value of $\chi^2$
as a function of $\Delta\alpha/\alpha$ -- should be near parabolic and
smooth. That is, $\Delta\chi^2_{\rm tol}$ should be $\ll 1$ to ensure
that fluctuations on the $\chi^2$ curve are also $\ll 1$.  This is
crucial for deriving the 1-$\sigma$ uncertainty in
$\Delta\alpha/\alpha$ from the width of the $\chi^2$ curve at
$\chi^2_{\rm min}+1$.

\begin{figure}
\includegraphics[width=\columnwidth]{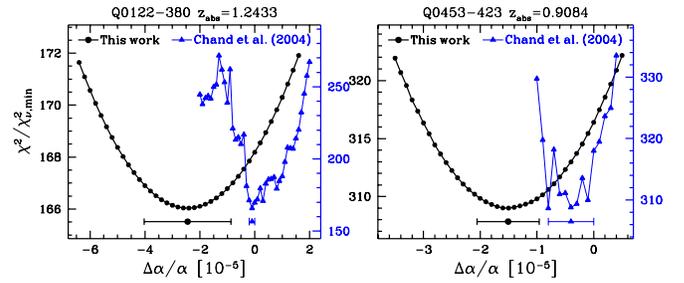}
\vspace{-2em}
\caption{Example $\chi^2$ curves from our minimization (circles) and
  that of \cite{SrianandR_04a} (triangles). Fluctuations in the latter
  indicate failings in the minimization. Points and error-bars
  indicate best-fitting values and 1-$\sigma$ uncertainties; for our
  curves $\Delta\alpha/\alpha$ was a free parameter. Note the
  different vertical scales: left-hand scales for our curves,
  right-hand scales for \cite{SrianandR_04a}.}
\label{fig:chi}
\end{figure}

However, none of Srianand {\it et al.}'s $\chi^2$ curves -- figure 2
in \cite{SrianandR_04a}, 14 in \cite{ChandH_04a} -- are smooth at the
$\ll 1$ level; many fluctuations exceed unity. Two examples are
reproduced in Fig.~\ref{fig:chi}. The fluctuations can only be due to
failings in the $\chi^2$ minimization: even when \cite{ChandH_04a} fit
\emph{simulated} spectra (their figure 2) jagged $\chi^2$ curves
result, leading to a strongly non-Gaussian distribution of
$\Delta\alpha/\alpha$ values and a large range of 1-$\sigma$
uncertainties (their figure 6). Clearly, these basic flaws in the
parameter estimation will yield underestimated uncertainties and
spurious $\Delta\alpha/\alpha$ values.

To demonstrate these failings, we apply the \emph{same profile fits}
to the \emph{same data} but with a robust $\chi^2$ minimization.
The spectra were kindly provided by B.~Aracil who confirmed that the
wavelength and flux arrays are identical to those in
\cite{SrianandR_04a}. For each absorber, the best-fitting profile
parameters of \cite{ChandH_04a} were treated as first guesses in our
$\chi^2$ minimization procedure (detailed in \cite{MurphyM_03a}).
The relationships between the Doppler widths of corresponding velocity
components in different transitions were also the same, as were the
relevant atomic data. The relative tolerance for halting the $\chi^2$
minimization was $\Delta\chi^2_{\rm tol}/\chi^2=2\times10^{-7}$. All
absorbers yield smooth $\chi^2$ curves in new our analysis;
Fig.~\ref{fig:chi} shows two examples.\vspace{6em}

By-products of this analysis are revised values of
$\Delta\alpha/\alpha$ and 1-$\sigma$ errors. We find 14 of the 23
$\Delta\alpha/\alpha$ values deviate by $>0.3\times10^{-5}$ from those
of \cite{SrianandR_04a}. Moreover, the errors are almost always
larger, typically by a factor of $\sim$3. The formal weighted mean
over the 23 absorbers becomes
$\Delta\alpha/\alpha=(-0.44\pm0.16)\times10^{-5}$ but the scatter in
the values is well beyond that expected from the errors. This probably
arises from many sources, including overly simplistic profile fits
(see \cite{MurphyM_07f}). Allowing for additional, unknown random
errors by increasing the error-bars to match the scatter
(i.e.~$\chi^2_\nu=1$ about the weighted mean), a more conservative
result from the data and fits of \cite{SrianandR_04a} is
$\Delta\alpha/\alpha=(-0.64\pm0.36)\times10^{-5}$ -- a 6-fold larger
uncertainty than quoted by \cite{SrianandR_04a}. We conclude that the
latter offers no stringent test of previous evidence for varying
$\alpha$; this must await a future, extensive statistical approach.

\bigskip
\noindent M.~T.~Murphy,$^{1,2}$ J.~K.~Webb,$^3$ V.~V.~Flambaum$^3$

\begin{small}
  $^1$Centre for Astrophysics \& Supercomputing, Swinburne

  $\phantom{^1}$University of Technology, Victoria 3122, Australia

  $^2$Institute of Astronomy, University of Cambridge
  
  $\phantom{^2}$Madingley Road, Cambridge CB3 0HA, UK

  $^3$School of Physics, University of New South Wales

  $\phantom{^3}$Sydney, NSW 2052, Australia

\end{small}


\newpage

\noindent \textbf{Discussion of Srianand \emph{et al.}'s Reply to our Comment}

\smallskip

Our \emph{Comment} (arXiv:0708.3677) sought to demonstrate that the
results of Srianand {\it et al.} \cite{SrianandR_04a} (also
\cite{ChandH_04a}) were not robust and were based on a measurement
technique which failed in a fundamental way. The numbers emerging from
the failed algorithm are meaningless, as discussed again below; they
cannot even be considered approximately correct. Indeed, when we apply
the same measurement technique (without the failure) to the \emph{same
  data}, using the \emph{same profile fits} as
\cite{SrianandR_04a,ChandH_04a}, we find very different values of
$\Delta\alpha/\alpha$ and errors which are typically a factor of
$\sim$3 larger. We present much more detail of that analysis in
\cite{MurphyM_07f}.

In their \emph{Reply} (arXiv:0711.1742), Srianand {\it et al.} state
or argue several points, all of which we dismiss below either because
they are demonstrably incorrect or because they rely on a flawed
application of simple statistical arguments. In order of importance:

\textbf{Point (i):} Despite demonstrating in our \emph{Comment} that
the measurement procedure used by \cite{SrianandR_04a,ChandH_04a}
failed, their \emph{Reply} argued that indeed their ``procedure is
robust as shown in'' \cite{ChandH_04a}. Much of the discussion and
Fig.~\ref{fig:chi} in our \emph{Comment} -- and, indeed, Srianand {\it
  et al.}'s own figures (2 in \cite{SrianandR_04a}, 14 in
\cite{ChandH_04a}) -- demonstrate the precise opposite, i.e.~that
large fluctuations on the $\chi^2$ curves are clearly present. This
means that at any given point on a $\chi^2$ curve (i.e.~for a given
input value of $\Delta\alpha/\alpha$), the true minimum of $\chi^2$
cannot have been reached. Therefore, the \emph{true} $\chi^2$ curve
must lie entirely \emph{beneath} that derived by
\cite{SrianandR_04a,ChandH_04a} for most absorbers. That is to say,
the $\chi^2$ curves of \cite{SrianandR_04a,ChandH_04a} simply cannot
be used to infer any values of $\Delta\alpha/\alpha$ or their
1-$\sigma$ errors whatsoever. Nor can values of $\Delta\alpha/\alpha$
(or uncertainties) one chooses to infer from them even be considered
`approximately correct' in any meaningful way. These basic and
fundamental aspects of $\chi^2$ fitting cannot be overemphasized:
\textbf{\boldmath{$\chi^2$} curves with large fluctuations provide no
  meaningful measurement of any kind}.

In their \emph{Reply}, Srianand {\it et al.} also ``point out that
fluctuations in $\chi^2$ curves get indeed smoothed after a large
number of iterations but the results from the first and last
iterations are found to be very similar''. It must again be strongly
emphasized that fluctuations on a $\chi^2$ curve indicate nothing but
the simple fact that $\chi^2$ has not been reduced to its true minimum
value at some, if not all points on the curve. One can not ``smooth''
these fluctuations in any way, not by averaging many ``iterations''
together (as may be implied by the above statement), nor by fitting a
parabolic or polynomial line through the $\chi^2$ curve as one would
fit a model to noisy data. It is not ``noise'' at all, but just an
indication that the algorithm for reducing $\chi^2$ has failed. One
simply has to identify the coding error, bug or mis-use of the
algorithm which is preventing $\chi^2$ from reaching its minimum.

Srianand \emph{et al.} \cite{SrianandR_04a,ChandH_04a} also argue that
their measurement procedure is robust based on measurements using
simulated absorption systems and these are referred to again in their
\emph{Reply}. However, as we pointed out in our \emph{Comment} (and in
more detail in \cite{MurphyM_07f}), those simulations actually
demonstrate the precise opposite: strong fluctuations even appear in
the $\chi^2$ curves for these simulations (figure 2 in
\cite{ChandH_04a}). This leads to spurious $\Delta\alpha/\alpha$
values: figure 6 in \cite{ChandH_04a} shows the results from 30
realizations of a simulated single-component Mg/Fe{\sc \,ii} absorber.
At least 15 $\Delta\alpha/\alpha$ values deviate by $\ge1\sigma$ from
the input value; 8 of these deviate by $\ge2\sigma$ and 4 by
$\ge3\sigma$. There is even a $\approx5$-$\sigma$ value. The
distribution of $\Delta\alpha/\alpha$ values should be Gaussian in
this case but these outliers demonstrate that it obviously is not. The
$\chi^2$ fluctuations also cause the uncertainty estimates from the
different realizations to range over a factor of $\approx4$ even
though all had the same simulated spectral signal-to-noise and input
profile fitting parameters.  None of these problems arise in our own
simulations of either single- or multiple-component systems (see
\cite{MurphyM_03a} for detailed discussion).

\textbf{Point (ii):} We used smaller error spectra in our
\emph{Comment} compared to the original analysis of
\cite{SrianandR_04a,ChandH_04a}, as described in detail in
\cite{MurphyM_07f}. Although Srianand {\it et al.} point out this fact
in their \emph{Reply}, they do not discuss its import. We do so in
\cite{MurphyM_07f}. To summarize: The main argument in our
\emph{Comment} is that the error bars of
\cite{SrianandR_04a,ChandH_04a} are underestimated. The fact that we
used \emph{smaller} error arrays and still found much \emph{larger}
errors on $\Delta\alpha/\alpha$ than \cite{SrianandR_04a,ChandH_04a}
only emphasizes this argument more. Using somewhat larger error arrays
would simply lead to somewhat larger uncertainties on
$\Delta\alpha/\alpha$ and change the actual measured values of
$\Delta\alpha/\alpha$ negligibly. Thus, ``Point 2'' in the \emph{Reply}
actually reinforces our conclusion that the data and fits do not offer
a stringent test of previous evidence for a varying $\alpha$.

\textbf{Point (iii):} The \emph{Reply} argues that many of our revised
values of $\Delta\alpha/\alpha$ ``match'' the original values of
\cite{ChandH_04a} ``at $\le 1\sigma$ level'' and, therefore, that the
original results are robust. Three simple points can be made here; the
first two (1 \& 2) are practical while the third (3) is a more
important general one:
\begin{enumerate}\vspace{-0.5em}
\item A matching criterion of ``$\le 1\sigma$'' is used for each value
  of $\Delta\alpha/\alpha$, but it isn't clear if ``$\sigma$'' is the
  original uncertaintiy from \cite{SrianandR_04a,ChandH_04a} or our
  revised value which is typically larger by a factor of $\sim$3. More
  importantly, we cannot replicate the number of systems which
  supposedly ``match'', i.e. 16. Using the uncertainty values from
  \cite{SrianandR_04a,ChandH_04a} as ``$\sigma$'' we find 11
  `matching' values of $\Delta\alpha/\alpha$. Using our new
  uncertainties gives 13 `matches'. Finally, adding the two
  uncertainty values in quadrature and treating that as ``$\sigma$''
  gives 14 systems.\vspace{-0.5em}
\item Even if the old and new $\Delta\alpha/\alpha$ values were
  statistically independent, how many values would one expect to
  ``match'' given the general distribution of values and uncertainties
  in both measurement sets? Whatever that number, even more values
  would ``match'' when we consider that the new and old values are in
  fact correlated. Therefore, at least some of the 16 `matches' the
  \emph{Reply} cites cannot be argued to bolster a case for the
  robustness of either the old or new values. Only if one found many
  more matches than one expects, given the two sets of values and
  their errors, would there be some case for that.\vspace{-0.5em}
\item The more important general point is that the usual ``1
  $\sigma$'' threshold is \emph{meaningless} here because the new and
  old $\Delta\alpha/\alpha$ values are not independent in any respect.
  When such non-independent data are being compared, it should
  actually be \emph{very few values} (not 32\% as for truly
  independent values) which would deviate by more than one formal
  standard deviation. Only if the deviation was very small with
  respect to the (old or new) 1-$\sigma$ uncertainties could one claim
  that the faults in the $\chi^2$ minimization algorithm had
  negligible importance and that the original results were robust. The
  large deviations we observe -- 14 of the new 23
  $\Delta\alpha/\alpha$ values deviate from the old by $> 0.3\times10^{-5}$
  -- clearly imply that the old results were not robust at all.
\end{enumerate}

\textbf{Point (iv):} The \emph{Reply} discusses how two
$\Delta\alpha/\alpha$ values in our revised set deviate by more than 4
$\sigma$ (presumably with respect to $\Delta\alpha/\alpha=0$) and
Srianand {\it et al.} choose to remove them from the sample. How can
one decide which systems are best to remove based only on the
parameter of interest, in this case the value (or significance) of
$\Delta\alpha/\alpha$? This is obviously a very biased selection
method. In our \emph{Comment}, this is why we chose to increase the
error bars on all $\Delta\alpha/\alpha$ values to match the observed
scatter.  Although itself not ideal, this is at least an unbiased
procedure. Our main point remains whatever procedure one chooses: the
constraints obtained from the data and fits of
\cite{SrianandR_04a,ChandH_04a} can not be regarded as robust, nor are
they a stringent test of previous evidence for varying $\alpha$.

\textbf{Point (v):} Srianand {\it et al.}'s \emph{Reply} states that
their ``procedure takes into account the differences in spectral
resolution in different settings ...~while this is not the case with
{\sc vpfit}''. This is simply incorrect, as is easily verified by
reading the {\sc vpfit} manual at
http://www.ast.cam.ac.uk/$\sim$rfc/vpfit.html [for the current
versions (8 \& 9), this is mentioned in the first few lines of the
introduction]. {\sc vpfit} has \emph{always} has this feature.
Differences in resolution between different portions of input spectra
were taken into account in all previous analyses of Keck/HIRES spectra
in \cite{WebbJ_99a,MurphyM_03a,MurphyM_04a} using {\sc vpfit}.

\end{document}